\documentclass[a4paper,fleqn]{cas-dc}

\usepackage[numbers]{natbib}
\usepackage{graphicx}
\usepackage{booktabs}
\usepackage{array}
\usepackage{ragged2e}
\usepackage{multirow}
\usepackage{tabularx}

\def\tsc#1{\csdef{#1}{\textsc{\lowercase{#1}}\xspace}}
\tsc{WGM}
\tsc{QE}
\tsc{EP}
\tsc{PMS}
\tsc{BEC}
\tsc{DE}

\begin{document}
\let\WriteBookmarks\relax
\def\floatpagepagefraction{1}
\def\textpagefraction{.001}
\shorttitle{Benchmarking IF Image Synthesis}
\shortauthors{X. Xing et~al.}

\title [mode = title]{Can Generative AI Replace Immunofluorescent Staining Processes? A Comparison Study of Synthetically Generated CellPainting Images from Brightfield}                      

\tnotetext[1]{This study was supported in part by the ERC IMI (101005122), the H2020 (952172), the MRC (MC$\slash$PC$\slash$21013), the Royal Society (IEC$\backslash$NSFC$\backslash$211235), the NVIDIA Academic Hardware Grant Program, the SABER project supported by Boehringer Ingelheim Ltd, Wellcome Leap Dynamic Resilience, and the UKRI Future Leaders Fellowship (MR$\slash$V023799$\slash$1).}


\author[1]{Xiaodan Xing}
\fnmark[1]
\affiliation[1]{organization={Bioengineering Department and Imperial-X, Imperial College London},
                city={London},
                country={United Kingdom}}

\author[1]{Siofra Murdoch} 
\fnmark[1]

\author[2]{Chunling Tang}
\fnmark[1]
\affiliation[2]{organization={Centre for Craniofacial \& Regenerative Biology, King’s College London},
                city={London},
                country={United Kingdom}}

\author[3,4]{Giorgos Papanastasiou}
\affiliation[3]{organization={Archimedes Unit, Athena Research Centre},
                city={Athens},
                country={Greece}}
\affiliation[4]{organization={School of Computer Science and Electronic Engineering, The University of Essex},
                city={Essex},
                country={United Kingdom}}                
\author[5]{Jan Cross-Zamirski}
\affiliation[5]{organization={Department of Applied Mathematics and Theoretical Physics, University of Cambridge},
                city={Cambridge},
                country={United Kingdom}} 

\author[2]{Yunzhe Guo}

\author[1]{Xianglu Xiao}
\affiliation[6]{organization={Data Sciences and Quantitative Biology, Discovery Sciences, AstraZeneca R\&D},
                city={Cambridge},
                country={United Kingdom}}

\author[5]{Carola-Bibiane Schönlieb}

\author[6]{Yinhai Wang}

\author[1,7,8,9]{Guang Yang}
\affiliation[7]{organization={National Heart and Lung Institute, Imperial College London},
                city={London},
                country={United Kingdom}} 
\affiliation[8]{organization={School of Biomedical Engineering \& Imaging Sciences, King's College London},
                city={London},
                country={United Kingdom}} 
\affiliation[9]{organization={Cardiovascular Research Centre, Royal Brompton Hospital},
                city={London},
                country={United Kingdom}} 

\cormark[1]
\ead{gyang@imperial.ac.uk}

\cortext[cor1]{Corresponding author}
\fntext[fn1]{Xiaodan, Siofra and Chunling contributed equally to this work.}


\begin{abstract}
Cell imaging assays utilizing fluorescence stains are essential for observing sub-cellular organelles and their responses to perturbations. Immunofluorescent staining process is routinely in labs, however the recent innovations in generative AI is challenging the idea of \textcolor{black}{wet lab immunofluorescence (IF) staining}. This is especially true when the availability and cost of specific fluorescence dyes is a problem to some labs. Furthermore, staining process takes time and leads to inter-intra-technician and hinders downstream image and data analysis, and the reusability of image data for other projects. Recent studies showed the use of generated synthetic IF images from brightfield (BF) images using generative AI algorithms in the literature. Therefore, in this study, we benchmark and compare five models from three types of IF generation backbones—CNN, GAN, and diffusion models—using a publicly available dataset.  This paper not only serves as a comparative study to determine the best-performing model but also proposes a comprehensive analysis pipeline for evaluating the efficacy of generators in IF image synthesis. We highlighted the potential of deep learning-based generators for IF image synthesis, while also discussed potential issues and future research directions. Although generative AI shows promise in simplifying cell phenotyping using only BF images with IF staining, further research and validations are needed to address the key challenges of model generalisability, batch effects, feature relevance and computational costs.

\end{abstract}

\begin{graphicalabstract}
\includegraphics[width=\textwidth]{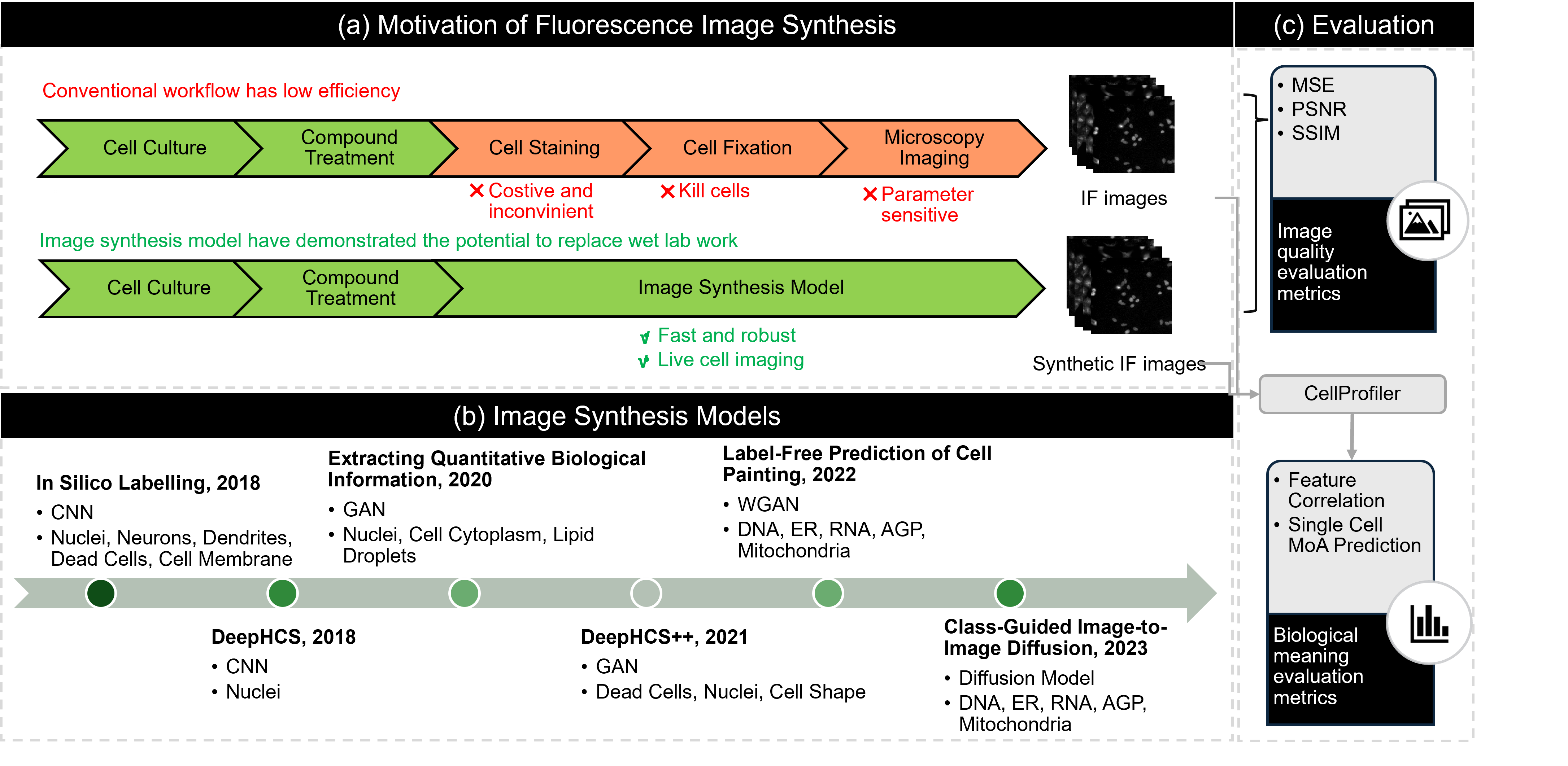}
\end{graphicalabstract}

\begin{highlights}
\item This study benchmarks five models from three types of IF generation backbones—CNN, GAN, and diffusion models—using a publicly available dataset, providing a detailed comparative analysis.

\item  Synthetic immunofluorescence images generated by deep learning models exhibit high visual quality and can effectively differentiate single-cell mechanism of action under various compound treatments. 

\item  The generalisability of the synthesis models is a concern. When tested on unseen cell lines, performance dropped sharply. Additionally, despite high mechanisms of action prediction accuracy, most feature correlations are low, raising concerns about the trustworthiness of these algorithms.

\end{highlights}


\maketitle

\section{Introduction}
High-throughput screening (HTS) has emerged as a critical method in drug discovery, enabling the fast evaluation of vast libraries of compounds for their therapeutic potential. In this context, cell image assays that utilise fluorescence microscopy are of great importance, as they allow scientists to observe and analyse cellular responses to pharmaceutical compounds. By applying fluorescent markers to highlight specific cellular structures, these assays yield detailed information about the effects of drugs on cell shape and function.

However, HTS via cell imaging assays presents distinct challenges, particularly in data acquisition and analysis. The first issue is the resource efficiency. Fluorescence microscopy requires specific dyes or labels, which can be expensive or limited in availability. In labs, certain fluorescent stains are often not immediately available, which inconveniently delays experiments. Additionally, the cost of fluorescence painting is also a concern. Specific dyes, such as the nucleic acid dye TOTO-3 iodide that has strong binding affinity for dsDNA, are very expensive, hindering research feasibility and scale especially for smaller labs. Finally, some specific fluorescent dyes introduce cytotoxicity to live cells and alter endogenous cell function and phenotype. Synthesising fluorescent images negates this issue, allowing for non-destructive analysis of cells and detection of subtle organelle signals. Brightfield (BF) imaging uses light and stains to visualise general tissue morphology and cell structure. Although easier to acquire,  the images are insufficient for detailed cellular and molecular analysis

Generative AI algorithms steps in. In 2018, Christiansen et al. \cite{christiansen2018silico} first proposed the use of CNN to predict a range of biological labels, such as those for nuclei, cell type (e.g., neural), and cell state (e.g., cell death). Notably, their results showed a high correlation between the location and intensity of the actual and predicted IF images, demonstrating that BF images contain sufficient information to train deep neural networks to predict IF staining and, therefore, show great promise as a solution to the time consuming, resource intensive, and toxic cell painting procedure. The potential of generative AI was further evidenced by a series of follow up studies \cite{lee2018deephcs,lee2021deephcs++,helgadottir2021extracting,cross2022label,cross2023class} that conducted experiments on a variety of other cell types and IF stains \cite{lamiable2023revealing,bourou2023phendiff}. 

\begin{figure*}[ht]
\centering
\includegraphics[width=\linewidth]{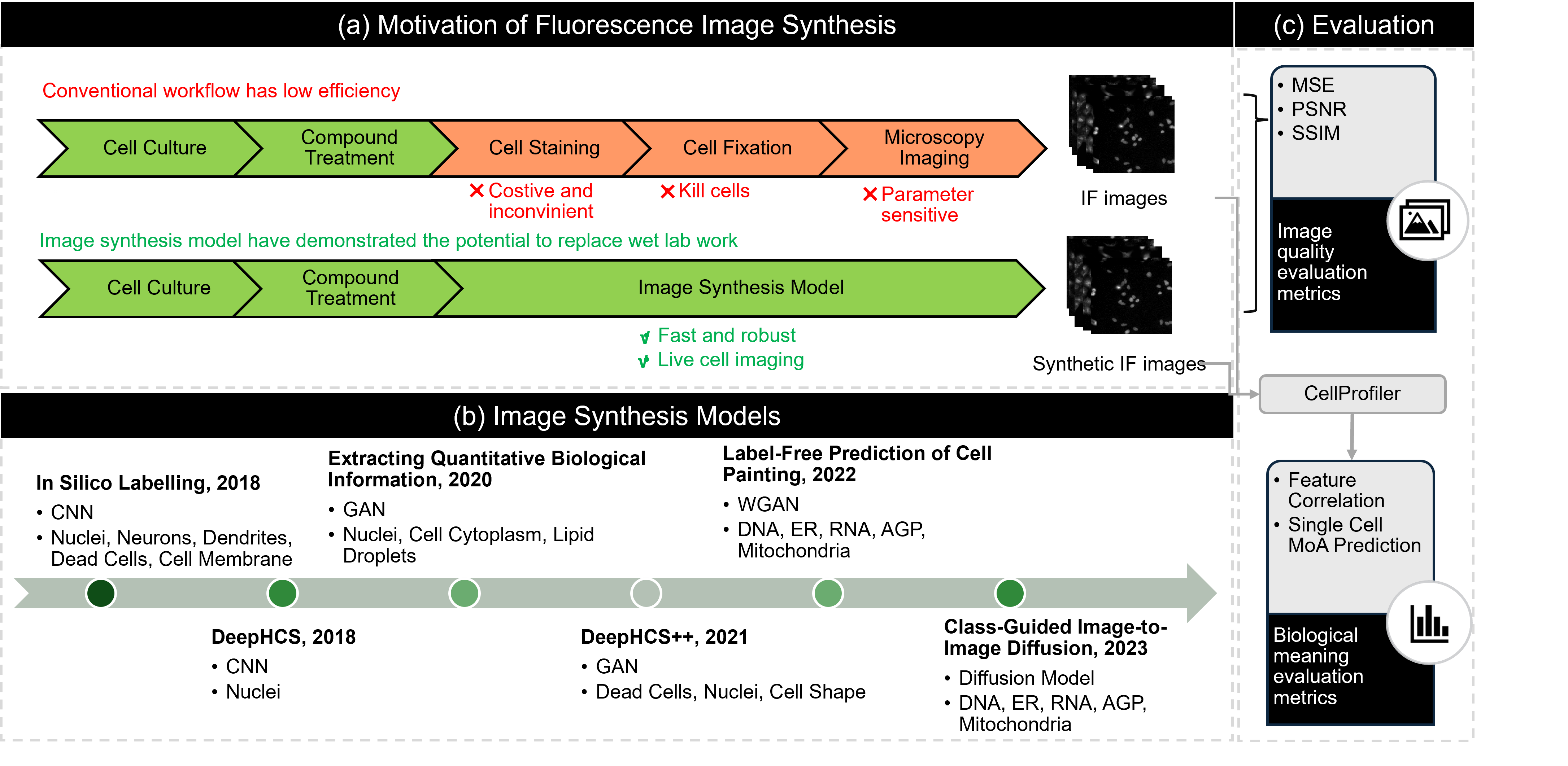}
\caption{Motivation of IF image synthesis from BF images (a), current IF image synthesis models (b) and evaluation metrics (c). }
\label{fig:intro}
\end{figure*}

However, a comprehensive review of these methods is lacking. Additionally, the datasets and image resolutions used in these studies vary, preventing an easy comparison between methods. Moreover, the narrowed focus on specific IF stains by most methodologies raises questions about generalisability and adaptability. Lastly, there is no common analysis pipeline for these methods, and the quality evaluation systems for synthetic IF images are inconsistent across previous studies.

Therefore, this project aims to benchmark selected IF generation methods using a publicly available dataset and provide guidance for model selection in IF image synthesis. We implemented five neural network frameworks from three popular generative genres, trained and evaluated more than 30 deep learning models, utilising over 1500 GPU hours. We have made all our training data, source codes, and checkpoints publicly available.

We aim to establish a foundation for the broader application of generative AI using cell imaging assays in preclinical settings. Our work showcases the potential of synthetic data in improving the efficiency of cell phenotypic analysis using only brightfield images. By addressing current research gaps, our project strives to set a new standard for using AI to predict IF staining. Ultimately, this work could significantly accelerate and streamline workflows in biomedical research. Finally, we conclude this research by answering the question that GenAI cannot entirely replace IF staining at present.

\section{Related Work in the Synthesis of Pseudo-IF Images }
In deep learning, the task of predicting IF images from BF images is posed as an image translation problem that aims to convert an image from one domain to another while preserving the original content. Figure \ref{fig:intro} summarises the backbones of current IF image synthesis algorithms and their target IF channels. 


Christiansen et al. \cite{christiansen2018silico} used convolutional neural networks (CNNs) optimised via Mean Squared Error (MSE) loss to minimise the difference between predicted IF images and actual stained IF images. Building on this, Lee et al. \cite{lee2018deephcs} enhanced the synthesis process with a course-to-fine approach, leveraging a Transformation Network to convert BF images to IF images and a Refinement Network for detail improvements. While effective, such CNN models produce blurred images due to the averaging effect of MSE loss, which smoothed out important details.

In 2014, the introduction of Generative Adversarial Networks (GANs) \cite{goodfellow2020generative} marked a significant improvement in image translation tasks. A GAN is composed of two main components: a generator creating synthetic images and a discriminator which evaluates whether an image is real or generated. The two networks are trained adversarially: while the generator strives to produce more realistic images, the discriminator improves its ability to distinguish between real and generated images. In the context of IF image synthesis, researchers \cite{helgadottir2021extracting,lee2021deephcs++,cross2022label} compared the performance of traditional CNNs and GAN models, demonstrating the superior performance of GANs in this task.

More recently, diffusion models \cite{ho2020denoising} have emerged as a novel technology in image synthesis. These models iteratively add noise to images and then reverse this process to restore the original images with high fidelity. Diffusion models outperform GANs \cite{dhariwal2021diffusion} in terms of fidelity and diversity due to their ability to preserve intricate details. Their effectiveness specifically in synthesizing IF images was validated in \cite{cross2023class}. 

Despite such technological advancements, standardised comparisons between diffusion models and GANs are lacking. Additionally, most studies focus on specific IF channels, as indicated in Table 1, resulting in a fragmented understanding of each model's strengths across different cellular structures. This gap emphasises the need for comprehensive benchmarks that assess CNNs, GANs, and diffusion models in a unified framework. Moreover, the selective focus on specific IF channels raises questions about the generalisability of these models. While some excel at depicting nuclear components, their ability to capture other structures, such as mitochondria or the cytoskeleton, may not be as robust. A holistic evaluation across various IF channels would yield more useful insights, leading to the development of more versatile image translation technologies that have practical use in labs.



\section{Generative Methods and Evaluations}
To bridge the research gap in the comprehensive analysis of various methods for IF image synthesis, this study aims to establish a standardised framework for comparative analysis by unifying test datasets, performance metrics, and evaluation protocols. We benchmarked three major types of image-to-image generation AI methods — CNNs, GAN-based models, and diffusion models — to deliver a thorough evaluation of the feasibility of generative AI in IF image synthesis. In this section, we offer a detailed overview of the dataset used, the models compared along with their experimental settings, and the evaluation metrics measured.

\subsection{Dataset}
Our model was trained on the public cpg0000 dataset \cite{chandrasekaran2024three} from the Cell Painting Gallery \footnote{\url{https://registry.opendata.aws/cellpainting-gallery/}}, which consists of over three million pairs of BF and IF images. This dataset includes samples treated with over 300 compounds, 160 CRISPR gene knockouts in both A549 and U2OS cell backgrounds at two time points. To evaluate the generalisability of our model, we selected a subset of these images, focusing on eight plates exposed to various compounds, as detailed in Table \ref{table:dataset_details}. Each plate contains 384 wells, with each well either treated with a specific compound or DMSO as a control. Two plates were used for training, one for validation and one for testing. Additionally, two plates under different pertubations were added for evaluating the generalisability of generative models. 

\begin{table}[h]
\centering
\begin{tabular}{p{1.6cm}p{1.5cm}p{1cm}p{1cm}p{1cm}}
\hline
\textbf{Assay Plate Barcode} & \textbf{Perturbation} & \textbf{Cell Type} & \textbf{Number of Images} & \textbf{Purpose} \\ \hline
BR00116991 & compound & A549 & 27648 & Train \\ \hline
BR00116992 & compound & A549 & 27640 & Train \\ \hline
BR00116995 & compound & U2OS & 27648 & Train \\ \hline
BR00117024 & compound & U2OS & 27648 & Train \\ \hline
BR00116993 & compound & A549 & 27352 & Validate \\ \hline
BR00117025 & compound & U2OS & 27648 & Validate \\ \hline
BR00116994 & compound & A549 & 27576 & Test \\ \hline
BR00117026 & compound & U2OS & 27648 & Test \\ \hline
BR00118041* & crispr & A549 & 27560 & Test \\ \hline
BR00118045* & crispr & U2OS & 27648 & Test \\ \hline
\end{tabular}
\caption{Details of the dataset including assay plate barcodes, perturbations, cell types, number of images, and their purposes in the generative model development. *These two plates were used only in generalisability evaluation. }
\label{table:dataset_details}
\end{table}

Overall, eight channels were scanned in this dataset, including three channel BF images and five channel IF images. \textbf{BF Images}: The primary brightfield channel records images at the focal plane which coincides with the lowest fluorescence emission. This channel serves as a reference point for the subsequent elevated and lowered focal planes. A BFHigh channel captures images 5 µm above the primary brightfield focal plane and a BFLow channel provides imagery from 5 µm below. \textbf{IF Images}: The dataset includes five channels under the Cell Painting (CP) protocol each tagged with specific dyes to highlight various cellular organelles and components, including AGP (actin, Golgi, plasma membrane); DNA (nucleus); ER (endoplasmic reticulum); mito (mitochondria); and RNA (nucleoli and cytoplasmic RNA). 

The size of each square image is 645 $\mu m$ (1080 pixels in both height and width). During training and inference, we resized all images to $512\times 512$  pixels. \textcolor{black}{We enhanced the contrast by cutting off the peak and bottom 5\% of pixels in the histogram. All images were saved in the 0-255 range during evaluation and normalized to the 0-1 range during training. }

\subsection{Synthesis Models Compared in This Study}
In this section we provide a more detailed view of each of the synthesis models compared in this paper. Overall, we choose five models from three different generative AI genres. 

\subsubsection{CNN Model}
For the CNN model tasked with synthesizing IF images from BF images, we have chosen the UNet architecture \cite{ronneberger2015u}. The UNet model is characterized by its encoder-decoder structure: the encoder compresses the input image into a feature-rich, compact representation while the decoder then progressively reconstructs the target image from the encoded representation. UNets are particularly noted for their skip connections, which bypass the encoding layers, feeding information directly to the corresponding decoding layers. These connections are essential as they carry fine-grained spatial information across the network, aiding in the precise localisation needed for accurate synthesis. Mean Squared Error (MSE) loss was utilised to train the UNet, which measures the average squared difference between the estimated values and what is estimated.

\subsubsection{GAN Models}
\textcolor{black}{GANs consist of a Generator \( G \) that generates fake data from random noise \( z \) and a Discriminator \( D \) that distinguishes real from fake data. They are trained simultaneously in a minimax game where \( G \) tries to minimise \( \log(1 - D(G(z))) \) and \( D \) tries to maximise \( \log D(x) + \log(1 - D(G(z))) \). This adversarial process results in \( G \) producing data increasingly similar to the real data.} We have implemented two variations of GANs differentiated by how they incorporate the input BF image condition:

\textbf{Pix2pix Model \cite{isola2017pix2pix}.} In the Pix2Pix model, the input consists of BF images, and the output is the corresponding synthetic IF images. This model uses a conditional GAN framework. Unlike traditional GANs that merely discriminate between real and fake outputs, the Pix2Pix discriminator evaluates pairs of condition (input BF image) and synthesised output (IF image), enhancing the relevance of generated images to the input conditions.

\textbf{SPADE \cite{park2019spade}.} The Spatially-Adaptive Denormalisation (SPADE) technique is employed to modify the normalisation process in the generator, making it adaptive to the input image condition. In our implementation, SPADE replaces traditional normalization blocks in the Pix2Pix architecture, enabling spatial information from the BF images is utilized in generating IF images.

\subsubsection{Diffusion Models}
\textcolor{black}{Diffusion models simulate the gradual transformation of data from a structured state into a random state and then learn to reverse this process to generate new data samples. The mathematical framework of diffusion models involves two key processes: the forward diffusion process, which adds noise to the data, and the reverse process, which learns to generate data by removing noise.}

\textcolor{black}{The forward process is modeled as a Markov chain that starts with the original data \( x_0 \) and adds noise at each step \( t \), resulting in a sequence of increasingly noisy data \( x_1, x_2, \ldots, x_T \), where \( T \) is the total number of diffusion steps. Mathematically, the transition from \( x_{t-1} \) to \( x_t \) can be defined using a Gaussian distribution:}

\[
x_t = \sqrt{1 - \beta_t} \, x_{t-1} + \sqrt{\beta_t} \, \epsilon_t,
\]

\textcolor{black}{where \( \beta_t \) is a variance schedule that determines the amount of noise added at each step, and \( \epsilon_t \sim \mathcal{N}(0, I) \) is a sample from a standard Gaussian distribution.}

\textcolor{black}{The reverse process is defined by a sequence of conditional distributions, each aiming to predict the clean data \( x_{t-1} \) given the noisy data \( x_t \):}

\[
p_\theta (x_{t-1} | x_t) = \mathcal{N}(x_{t-1}; \mu_\theta (x_t, t), \sigma_\theta (x_t, t)),
\]

\textcolor{black}{where \( \mu_\theta (x_t, t) \) and \( \sigma_\theta (x_t, t) \) are functions parameterized by the model parameters \( \theta \), predicting the mean and variance of the Gaussian distribution for \( x_{t-1} \) based on \( x_t \).}

\textbf{Palette \cite{saharia2022palette}.} Within the diffusion model framework, Palette utilises an attention U-Net architecture for the denoising process; the input BF images are concatenated with noisy images, providing conditional context throughout the iterative denoising steps. As with the UNet implementation, MSE loss is used to guide the denoising process towards producing high-quality IF images.

\textbf{SPADE-Diffusion \cite{wang2022spadediffusion}.} Combining the capabilities of SPADE for spatial adaptation with the robust image synthesis of diffusion models, SPADE-Diffusion represents a hybrid approach. This model leverages the strengths of both techniques to enhance the quality and relevance of the synthesised IF images, particularly in retaining and utilising spatial and contextual information from the BF inputs.

\subsection{Evaluation Methods}

Various algorithms have been proposed in previous studies to evaluate synthetic images \cite{heusel2017gans,salimans2016improved,borji2019pros}. In this paper, we define two categories of metrics: image quality evaluation metrics and biological meaning evaluation metrics. \textcolor{black}{Several images near the edges of the plates contained no useful information because the corresponding ground truth IF images for these regions were entirely black. To ensure more accurate analysis, we excluded these non-informative images from the dataset.}

\subsubsection{Image Quality Evaluation Metrics: MSE, PSNR and SSIM}
Given an image synthesis task that involves paired ground truth data, the most straightforward way to evaluate synthetic images is by determining the Mean Squared Error (MSE) between the predicted and real images. From a staining perspective, this method relevantly assesses the \textit{x} intensity difference between real and predicted IF images. Peak Signal-to-Noise Ratio (PSNR), which is computed based on MSR, similarly measures the error between synthetic and ground truth images. The Structural Similarity Index Measure (SSIM) evaluates how similar the structures in the synthetic image are to those in the ground truth. 

\textcolor{black}{Notably, within the field of image synthesis, many algorithms, such as FID  \cite{heusel2017gans} and Inception Score \cite{salimans2016improved}, evaluate the distribution difference between sets of real and synthetic image data rather than the specific differences between individual images.} However, due to the fact that we can obtain ground truth labelling and that the motivation behind this project is to assess corresponding correlation or predictive ability of BF images to IF images, such metrics, which are computed over the whole distribution, were not used. For a similar reason, image variety measurements \cite{xing2023beauty} were also not computed. 

\subsubsection{Biological Meaning Metrics: Feature Correlation and MoA Prediction Accuracy}
\begin{figure*}[h!]
\centering
\includegraphics[width=\linewidth]{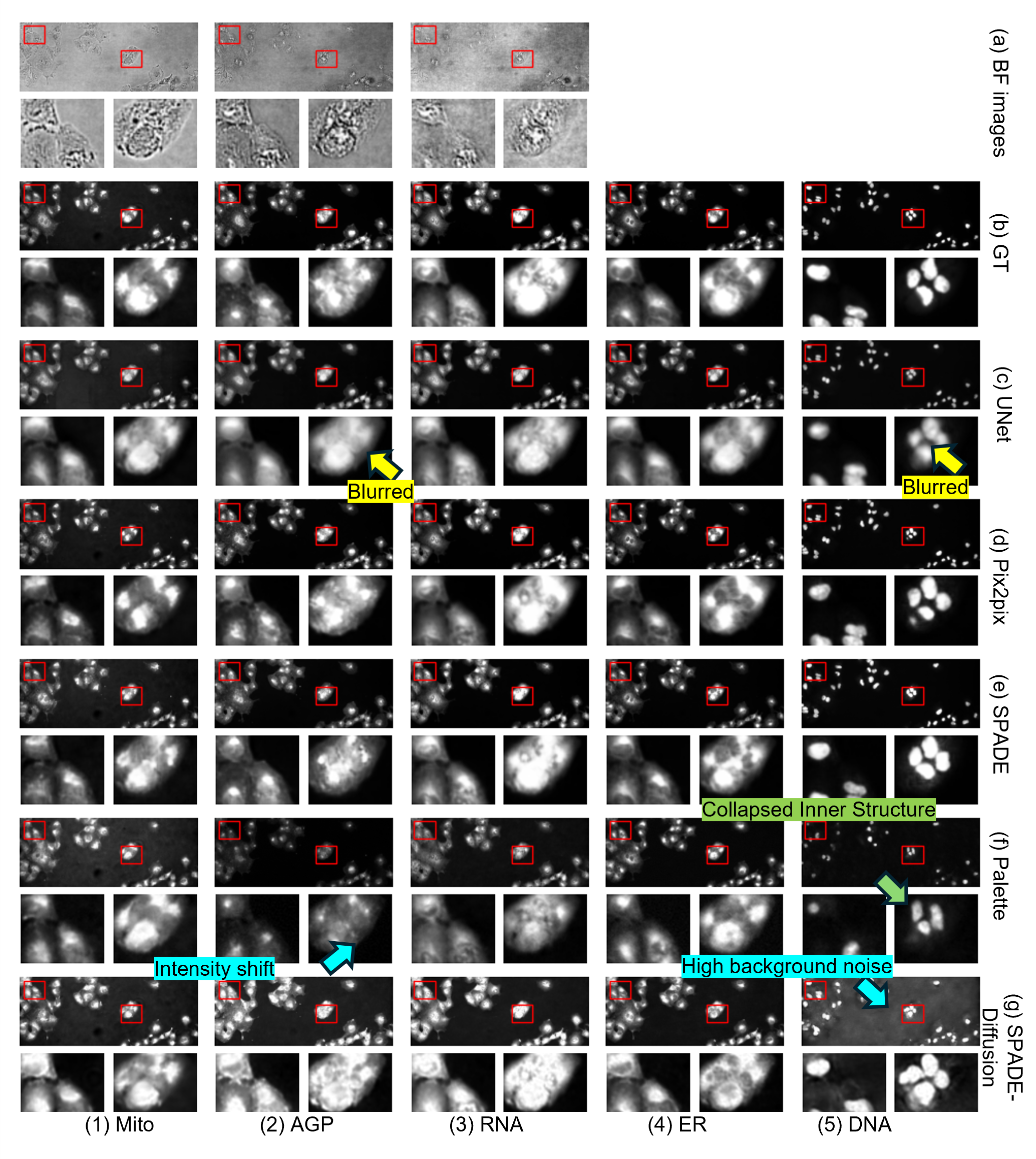}
\caption{Visualisation of synthesised image results. We noticed that the Pix2Pix model produces synthetic IF images with the highest visual quality compared to other methods. In the discussion section, we highlight and discuss three characteristic types of artifacts we identified when predicting IF images from BF images using generative AI.  }
\label{fig:results}
\end{figure*}

To assess biological meaning, we used CellProfiler bioimage analysis software (v4.0.6) to process the images adapted from the pipeline proposed in \cite{chandrasekaran2024three}. Following corrections for variations in background intensity, we then segment cells, distinguishing between compartments of cell, nucleus and cytoplasm levels. Then, across all five fluorescent channels, we measured cell features across several categories, including fluorescence intensity, texture, granularity and density. Following the image analysis pipeline\footnote{\url{https://github.com/jump-cellpainting/2024_Chandrasekaran_NatureMethods/tree/main/pipelines/2020_11_04_CPJUMP1}}, we obtained 1180 cell features, 1165 cytoplasm features and 1173 nuclei features.

The first evaluation metric is the correlation between these features. The mean values for each feature group were organized into correlation matrices and visualised as heatmaps. In CellProfiler, features from each object were divided into several categories: area/shape, colocalisation, granularity, intensity, neighbours, radial distribution, and texture.

The second evaluation metric is the predictive performance of these single-cell features for determining the mechanism of action (MoA) of perturbations e.g. compounds. The MoA refers to the biological process through which a compound achieves its pharmacological effect, such as targeting specific proteins or modulating certain pathways. Understanding a compound’s MoA is critical in pre-clinical phases as it helps identify potential efficacy and toxicity effects. Synthetic IF images can be used for automatic diagnosis \cite{zhao2022reasoning}, such as identifying diseased cells. However, we did not explore this due to the limited number of cell lines.

Cell morphological features extracted from IF images play a vital role in determining and confirming a compound’s MoA. To validate the efficiency of synthesised images, we trained two SVMs: one on features extracted from real IF images and the other on features extracted from synthetic IF images. Additionally, we categorised the inhibitor types into six groups based on their potential effects on IF images to reduce the number of classes in the classification task, as shown in Table \ref{tab:inhibitor_groups}. F1 scores for both the six-class classification and the binary (inactive cmpds vs. others) classification were reported using five-fold cross-validation. 

It should be noted that in this study the classification algorithm is trained and tested on the same type of dataset. For example, if an MoA prediction algorithm is trained on features extracted from the pix2pix model, we also report the test accuracy on features extracted from the pix2pix images. This approach differs from a "train-on-synthetic-test-on-real" setting, where the model trained on synthetic images would be tested on real images. The purpose of this accuracy reporting is to evaluate whether synthetic images are capable of demonstrating different MoAs effectively, rather than assessing the performance of the features themselves in training an accurate classifier.

\begin{table}[ht]
    \centering
    \begin{tabular}{>{\RaggedRight} p{3.5cm}>{\RaggedRight} p{4cm}}
        \hline
        \textbf{Group} & \textbf{Inhibitors} \\ \hline
        Group 1: Kinase Inhibitors & Aurora kinase inhibitor, Bcr-Abl kinase inhibitor, CDC inhibitor, DYRK inhibitor, EGFR inhibitor, JAK inhibitor, JNK inhibitor, p21 activated kinase inhibitor \\ \hline
        Group 2: Epigenetic Modifiers & HDAC inhibitor, bromodomain inhibitor \\ \hline
        Group 3: Growth Factor Receptor Inhibitors & IGF-1 inhibitor, hepatocyte growth factor receptor inhibitor \\ \hline
        Group 4: Protease Inhibitors & Ubiquitin specific protease inhibitor\\ \hline
        Group 5: Lipid Signaling Modifiers & Phospholipase inhibitor\\ \hline
        Control Group & DMSO (commonly used as neutual control in  experiments)\\ \hline
    \end{tabular}
    \caption{The chemical compounds are further categorized into six groups.
 }
    \label{tab:inhibitor_groups}
\end{table}

\section{Experimental Results}
\subsection{Pix2Pix Model Performs the Best Regarding Image Quality Evaluation}
\textbf{Visual Comparison. } \textcolor{black}{We invited two biologists who work closely with IF staining to visually compare the performance of different models. } Both agreed that through sufficient training, all five models captured the distinctive style of the IF images, though, there are noticeable differences in details between the generated images and the ground truth images. Among these models, we discovered that the Pix2Pix model captures the details of the real IF images the most accurately.

\begin{figure}[h!]
\centering
\includegraphics[width=\linewidth]{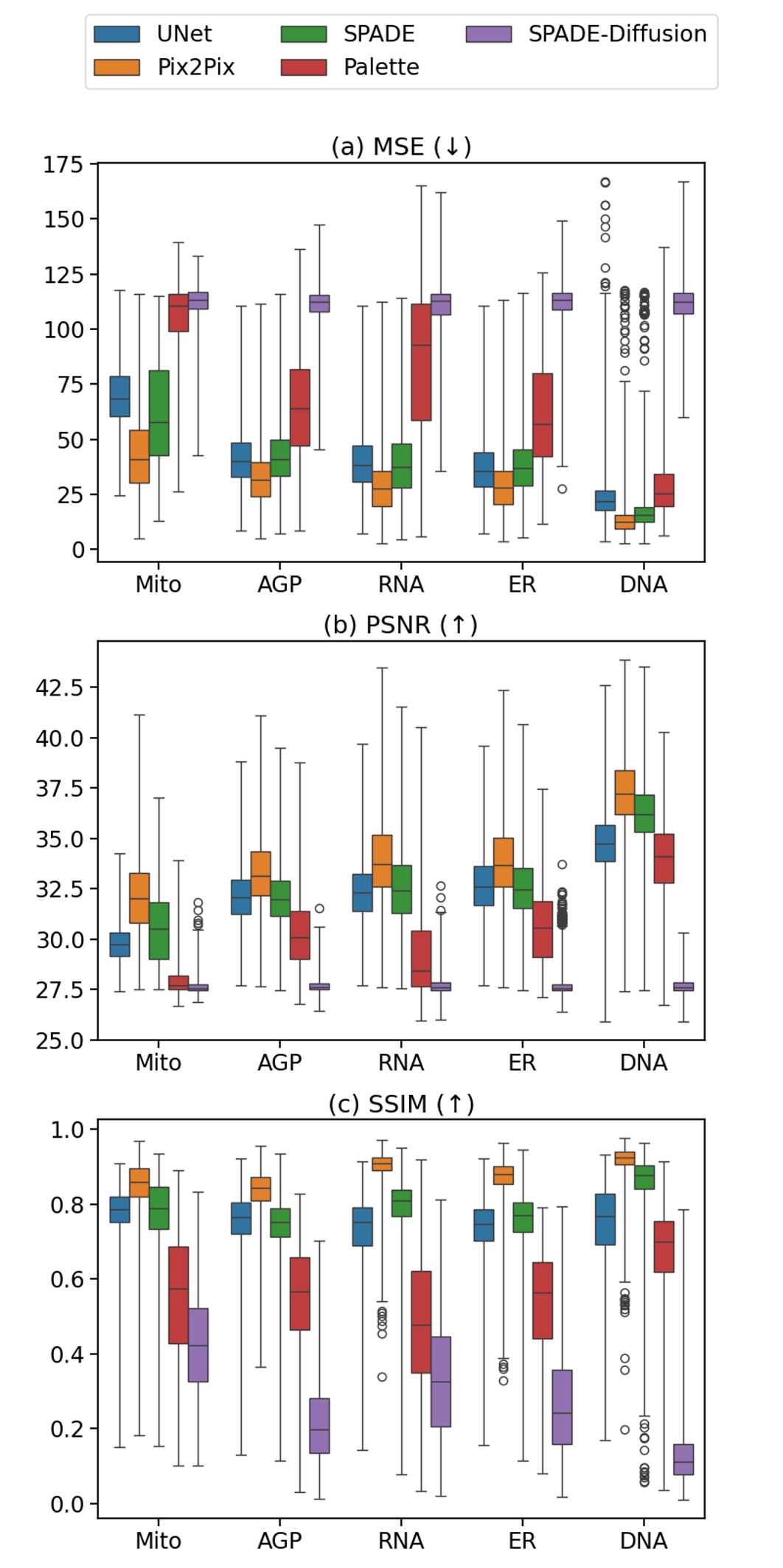}
\caption{Quantification results of image qualities from different models grouped by their channels numbers. Here, we present the image quality metrics MSE (a), PSNR (b) and SSIM (c) for all generative models, grouped according to channels 1 through 5, which correspond to Mito, AGP, RNA, ER, and DNA, respectively.}
\label{fig:mses}
\end{figure}

\textbf{Quantification Results. } We calculated the MSE, PSNR and SSIM for all images, and grouped them in Figure \ref{fig:mses}. The quantitative image quality evaluation further supported our claim that the Pix2Pix model is the best performing model across all channels in all metrics. 

In Figure \ref{fig:mses}(a), when comparing the MSE values across different channels, we noticed that, with the exception of SPADE, the MSEs for the DNA channel in all models are significantly lower ($p<0.05$) than those for other channels. This suggested that predicting the cell nucleus channel is easier to achieve than transferring the styles of BF or searching for subtle and embedded subcellular structures in other channels (Mito, AGP, ER, and RNA). This is likely because nuclei have a clearer shape and more defined semantics compared to other structures. To improve synthesis across all structures, additional conditional mechanisms may be required during the synthesis process across all model families.

\subsection{Biological Meaning Analysis}
Besides analyzing the imaging quality, we wish to further explore the biological meaning to understand whether the generative results can preserve important features compared to the original model and whether they can be used for other downstream applications.

\subsubsection{Most Features From Synthetic IF Images Have Low Correlations With Real Features}
\begin{figure}[ht]
\centering
\includegraphics[width=\linewidth]{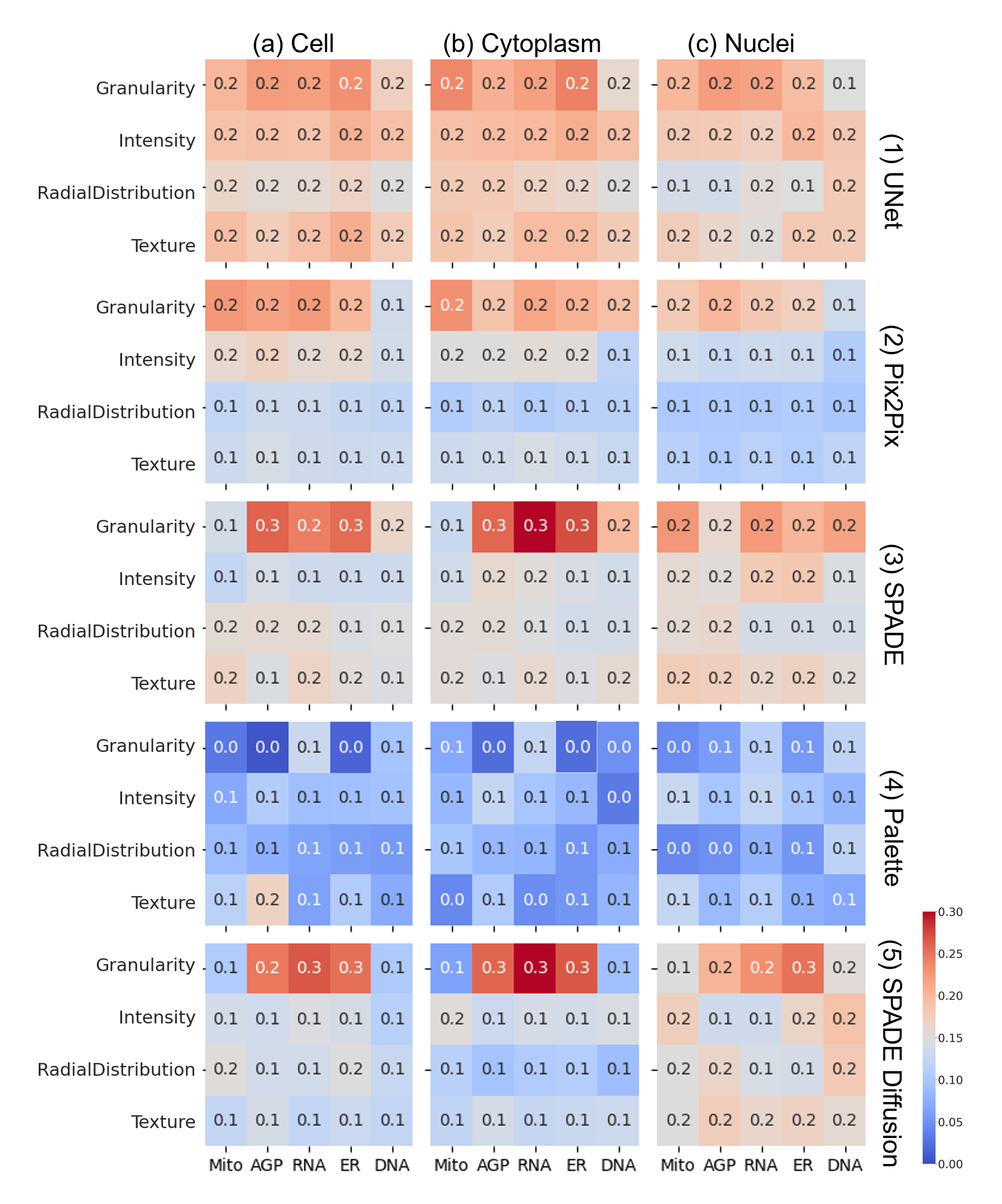}
\caption{Channel-wise Pearson's correlations between morphological features extracted from different synthetic methods and real features, aggregated by channel and feature group for cell, cytoplasm, and nuclei objects.}
\label{fig:cp_corr}
\end{figure}

\begin{figure}[ht]
\centering
\includegraphics[width=\linewidth]{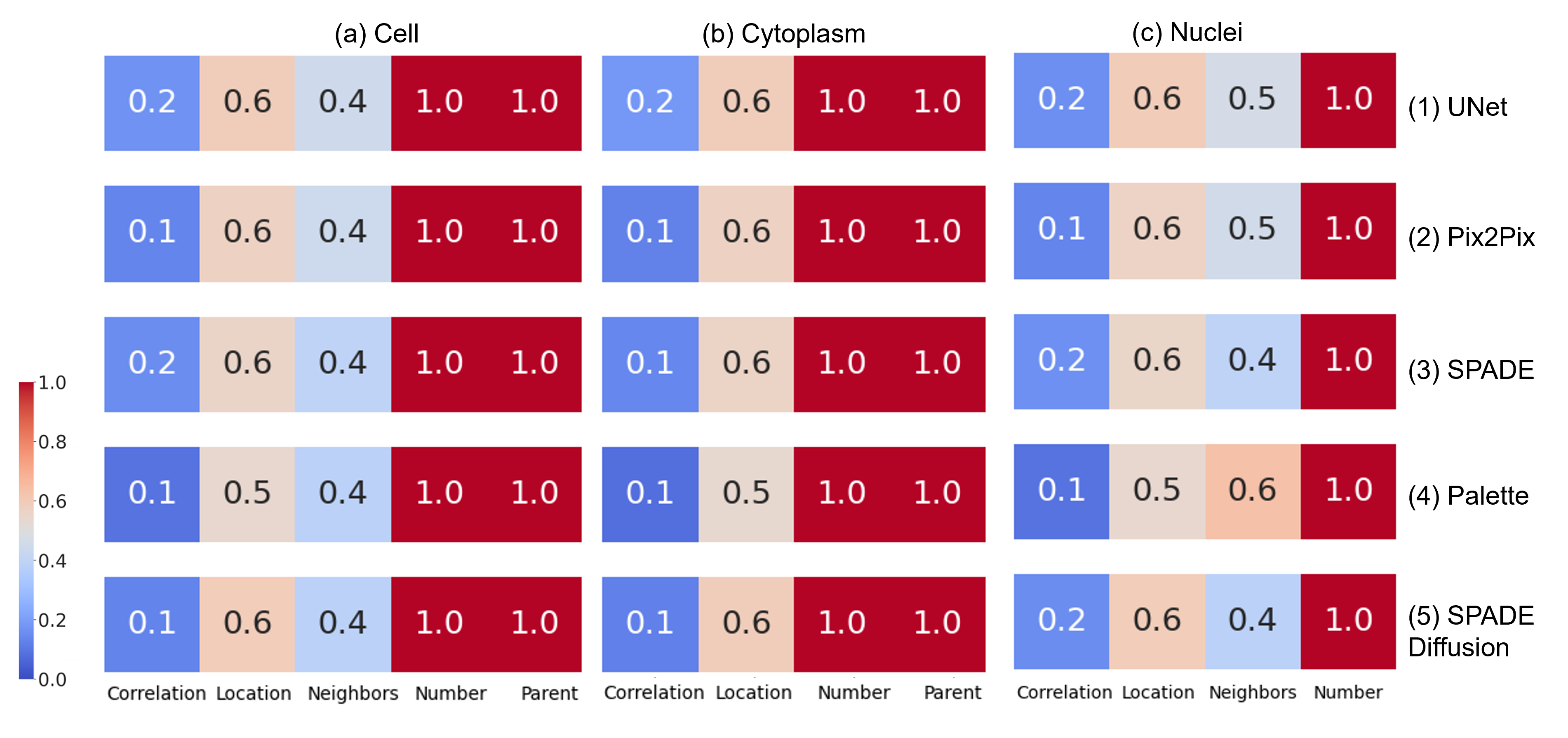}
\caption{Correlations between morphological features extracted from different synthetic methods and real features. These features are cell, cytoplasm, and nuclei objects.}
\label{fig:cp_corr2}
\end{figure}

Here we computed the correlation between hand-crafted features from CellProfiler analysis; results are shown in Figure \ref{fig:cp_corr} and Figure \ref{fig:cp_corr2}. We observed that most channel-wise features, i.e., features extracted from each channel as shown in Figure \ref{fig:cp_corr}, have a correlation lower than 0.3. In contrast, the overall mean correlation between real versus real channel-wise features was reported as 0.6 in \cite{cross2023class}. This significant difference highlights the disparity between biological features extracted from synthetic images and those from real images.

The correlation between the number of objects and the parent relation — where the feature "Parent" denotes the relationship between cells and corresponding sub-cellular structures — is nearly 1.0 in all models. This result is expected because such relationships can be accurately derived from BF images, enabling the synthesis model to capture this feature effectively.

We observed that most synthetic images produce granularity features with relatively high correlation with the real images. Granularity features in CellProfiler quantify the texture of objects within an image, specifically focusing on the size and distribution of fine subcellular structures (small particles or structures) within cells. Additionally, granularity features are regarded as very important in single-cell MoA classification tasks, specifically in synthetic images, as in Table \ref{table:results}. The high correlation of granularity features between synthetic and real images suggests that the synthetic models effectively replicate these critical texture patterns, which may contribute to the importance of these features in classification tasks.



\subsubsection{Most Features From Synthetic IF Images Predict MoA As Accurately As Real Features}
In Table \ref{table:results}, we show the F1-scores of the 5-fold validation results. This was done to evaluate if the features possess the classification ability to distinguish between DMSO and inhibitor treatment groups. We did not observe significant difference between classification results from the five models. All features can classify single-cell states from control to inhibitor treatment groups with approximately 70\% accuracy.

We also plotted the top five decisive features in the classifier. It was noted that in real images, the correlations between channels are crucial, while in synthetic images, granularity features are more decisive in the MoA prediction.

\begin{table*}[ht]
\centering
\scriptsize
\begin{tabularx}{\textwidth}{p{0.5cm} p{1cm} X X X}
\toprule
\textbf{Cell} & \textbf{Metric} & \textbf{GT} & \textbf{UNet} & \textbf{Pix2Pix} \\
\midrule
\multirow{3}{*}{A549} & F1-Score & 0.57 & 0.6 & 0.58 \\
 & F1-Score (DMSO) & 0.71 & 0.71 & 0.71 \\
 & Top 5 features & \parbox{1.5cm}{Correlation\_RWC\_DNA\_AGP\_Cells \\ Correlation\_RWC\_DNA\_RNA\_Cells \\ Correlation\_Correlation\_AGP\_DNA\_Cells \\ Correlation\_RWC\_DNA\_ER\_Cells \\ Granularity\_14\_Mito\_Cells} & \parbox{1.5cm}{Granularity\_16\_DNA\_Cytoplasm \\ Granularity\_16\_DNA\_Nuclei \\ Granularity\_16\_DNA\_Cells \\ Granularity\_15\_Mito\_Cells \\ Granularity\_16\_Mito\_Nuclei} & \parbox{1.5cm}{Granularity\_15\_DNA\_Nuclei \\ Granularity\_15\_DNA\_Cells \\ Granularity\_15\_DNA\_Cytoplasm \\ Granularity\_16\_Mito\_Cells \\ Granularity\_16\_ER\_Cytoplasm} \\
\midrule
\multirow{3}{*}{U2OS} & F1-Score & 0.58 & 0.56 & 0.57 \\
 & F1-Score (DMSO) &  0.70 & 0.69 &  0.70 \\
 & Top 5 features & \parbox{1.5cm}{Correlation\_RWC\_DNA\_AGP\_Cells \\ Correlation\_RWC\_DNA\_RNA\_Cells \\ Correlation\_Correlation\_AGP\_DNA\_Cells \\ Correlation\_RWC\_DNA\_ER\_Cells \\ Granularity\_14\_Mito\_Cells} & \parbox{1.5cm}{Correlation\_RWC\_ER\_AGP\_Cytoplasm \\ Correlation\_RWC\_RNA\_AGP\_Cytoplasm \\ Granularity\_14\_DNA\_Cells \\ Granularity\_15\_DNA\_Cells \\ Granularity\_14\_DNA\_Cytoplasm} & \parbox{1.5cm}{Granularity\_15\_Mito\_Cells \\ Granularity\_16\_Mito\_Nuclei \\ Number\_Object\_Number\_Cytoplasm \\ Granularity\_16\_Mito\_Cells \\ Correlation\_RWC\_DNA\_RNA\_Cells} \\
\bottomrule
\end{tabularx}

\end{table*}

\begin{table*}[ht]
\centering
\scriptsize
\begin{tabularx}{\textwidth}{p{0.5cm} p{1cm} X X X}
\toprule
\textbf{Cell} & \textbf{Metric} & \textbf{SPADE} & \textbf{Palette} & \textbf{SPADE Diffusion} \\
\midrule
\multirow{3}{*}{A549} & F1-Score & 0.58 & 0.69 &  0.70 \\
 & F1-Score (DMSO) &  0.70 & 0.77 & 0.77 \\
 & Top 5 features & \parbox{1.5cm}{Granularity\_16\_DNA\_Cells \\ Granularity\_16\_DNA\_Nuclei \\ Granularity\_15\_Mito\_Cytoplasm \\ Granularity\_16\_DNA\_Cytoplasm \\ Granularity\_16\_Mito\_Cells} & \parbox{1.5cm}{Granularity\_16\_DNA\_Nuclei \\ Granularity\_16\_DNA\_Cells \\ Granularity\_16\_DNA\_Cytoplasm \\ Granularity\_15\_DNA\_Cytoplasm \\ Granularity\_12\_DNA\_Nuclei} & \parbox{1.5cm}{Granularity\_15\_DNA\_Nuclei \\ Granularity\_13\_DNA\_Cells \\ Granularity\_15\_DNA\_Cells \\ Granularity\_15\_DNA\_Cytoplasm \\ Granularity\_16\_DNA\_Nuclei} \\
\midrule
\multirow{3}{*}{U2OS} & F1-Score & 0.55 & 0.71 & 0.69 \\
 & F1-Score (DMSO) & 0.69 & 0.77 & 0.77 \\
 & Top 5 features & \parbox{1.5cm}{Granularity\_16\_AGP\_Cells \\ Granularity\_13\_Mito\_Cells \\ Granularity\_13\_Mito\_Cytoplasm \\ Granularity\_16\_ER\_Cytoplasm \\ Granularity\_16\_DNA\_Cytoplasm} & \parbox{1.5cm}{Granularity\_15\_DNA\_Cytoplasm \\ Granularity\_16\_DNA\_Cytoplasm \\ Granularity\_1\_DNA\_Cells \\ Granularity\_15\_DNA\_Nuclei \\ Granularity\_16\_DNA\_Cells} & \parbox{1.5cm}{Granularity\_16\_DNA\_Cytoplasm \\ Granularity\_16\_AGP\_Nuclei \\ Granularity\_16\_DNA\_Nuclei \\ Granularity\_16\_DNA\_Cells \\ Granularity\_15\_DNA\_Nuclei} \\
\bottomrule
\end{tabularx}
\caption{MoA performance metrics and top 5 decisive features for A549 and U2OS cell lines using features extraced from real and generated IF images.}
\label{table:results}
\end{table*}


\subsection{Acceptable Time and Computational Cost of Synthetic IF Images Compared to Bench Work }
In this section, we provide information on the computational cost of the generators compared in this study. All models were trained and tested on NVIDIA RTX 6000 GPU clusters. We recorded the training hours for 10 epochs (optimising the entire training dataset 10 times) and the GPU memory usage during training for each model. For inference, we measured the time taken to process each image. We also recorded the file size of the model checkpoints for storage. It should be noted that the training time may vary in different working environments. 

The SPADE normalisation method increases computational requirements and GPU memory usage. Diffusion models have a significant bottleneck in inference time, as we used 1000 denoising steps for both models. \textcolor{black}{However, even the most time intensive computation model saves substantial time compared to real-world wet lab work. These lab staining tasks, which include processes such as fixation, blocking, and primary and secondary antibody incubation, typically takes 1-2 days of lab work. Therefore, when comparing the computational cost, we conclude that generative AI can significantly reduce the time required to obtain IF images.
}

In addition to the time cost, here we compared the financial costs related to both wet lab cell painting and virtual cell painting image synthesis. It is worth nothing that since the learning curves of researchers is both difficult and subjective to evaluate, we opt only to consider hardware and consumable -- not human -- resource costs in our comparisons.

\begin{table}[h]
\centering
\begin{tabular}{p{1.5cm}p{1cm}p{1.5cm}p{1cm}p{1.5cm}}
\hline
\textbf{Model} & \textbf{Train Time* (h)} & \textbf{GPU Memory** (GB)} & \textbf{Inference Time (s)} & \textbf{Checkpoint Memory (MB)} \\ \hline
UNet & 69h 34min & 3.31 & 0.06 & 134.98 \\ \hline
PIX2PIX & 10h 7min & 4.78 & 0.07 & 712.72 (21.64) \\ \hline
SPADE & 19h 30min & 6.50 & 0.10 & 360.14 (21.71) \\ \hline
PALETTE & 32h 33min & 6.24 & 40 & 98.29 \\ \hline
SPADE-DIFFUSION & 30h 17min & 17.79 & 240 & 657.85 \\ \hline
\end{tabular}
\caption{Computational costs of all models. *Training time for 10 epochs. **GPU memory for a batch size of 1 with 512x512 resolution. For wet lab work,  staining usually takes 1-2 days. }
\label{table:model_comparison}
\end{table}

\subsection{Generalisability to Unseen Circumstances}
Sections 4.1 and 4.2 illustrate the potential of generative AI to synthesise IF images due to the promising image quality and accuracy in MoA prediction. However, it is worth noting that models were both trained and tested on the same cell line and under identical compound treatments. While this confirms that the model can generate next-to-real cell morphology and generate accurate cell painting images when presented with familiar  data, it leaves open the question: is it possible to predict in unfamiliar scenarios such as new cells and unseen perturbations? 

\subsubsection{Lack of Generalisability to Unseen Cell Lines}
Our original implementation used two cell lines independently: A549 and U2OS. To investigate the generalisability of generative AI to new cell lines, we selected the best-performing Pix2Pix backbone and trained the generators using one cell line then tested on the other.

\begin{figure}[ht]
\centering
\includegraphics[width=\linewidth]{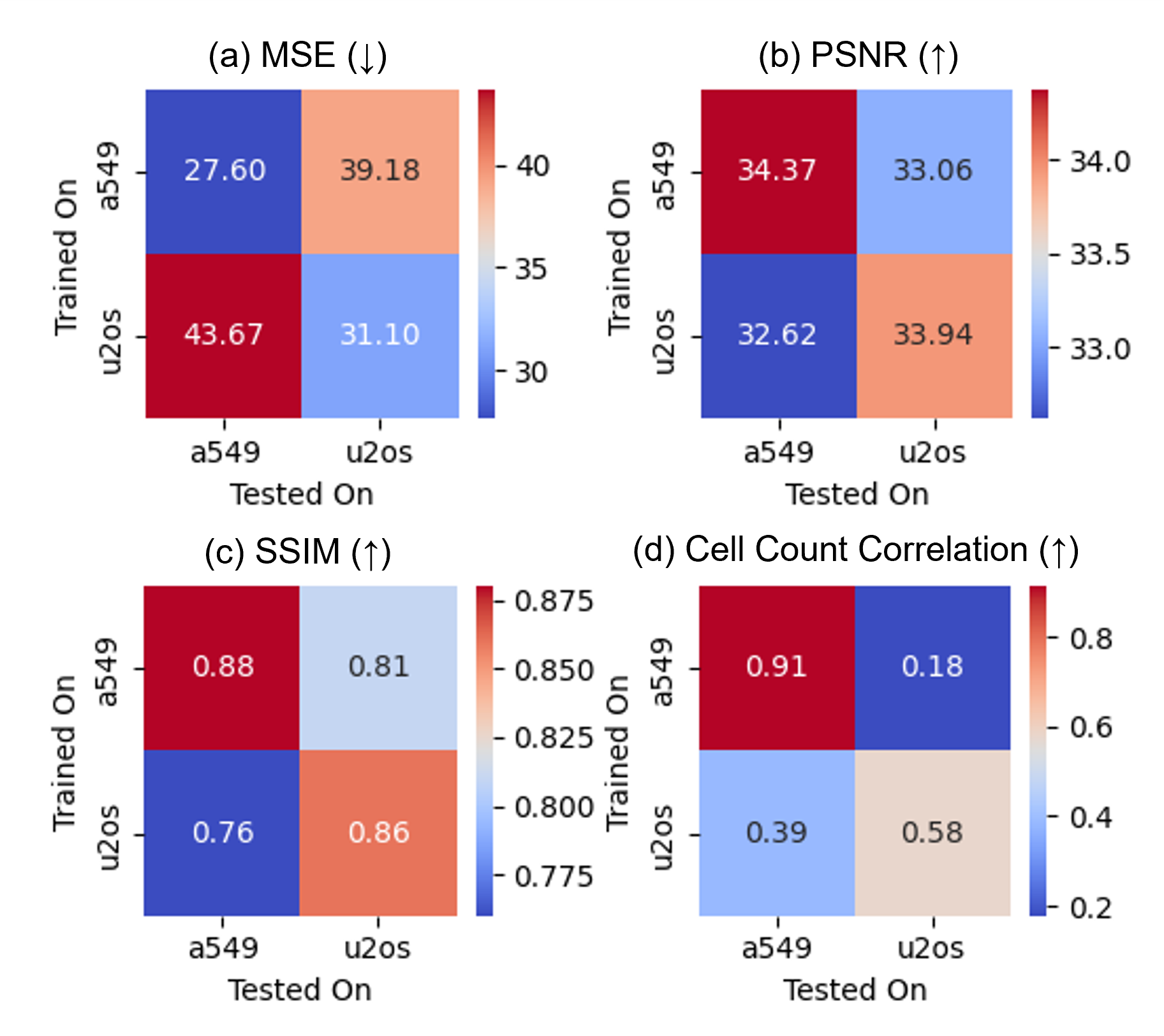}
\caption{The CellProfiler clustering results for cell phenotyping. Our synthetic image should have the same phenotyping clustering pattern with real images. }
\label{fig:cross}
\end{figure}

To evaluate model performance, we calculated the MSE, PSNR, and SSIM. For biological relevance, we tested the correlation of cell counts between the real and synthetic images. In Figure \ref{fig:cross}, we observed that the diagonal values in PSNR, SSIM and Cell were significantly higher ($p<0.05$) than the off-diagonal values (with MSE being lower), indicating a notable decline in performance when the models were used on unseen cell lines. This performance decline highlights the limited generalisability of deep generative AI to unseen cell lines, raising concerns about the method's practicality.

\subsubsection{Generalisability to Unseen Pertubations}
\textcolor{black}{In our experiments, we used microscopy images with compound treatments. However, gene editing can also perturb cells and alter their morphological features across different channels. To evaluate the generalisability of generative AI on unseen perturbations, we used our best-performing model, Pix2Pix, which was trained on BF-IF image pairs with compound treatments, and tested it on gene-edited A549 and U2OS cell lines.}

\textcolor{black}{Although we discovered a significant performance drop in synthetic image quality (MSE 37.57, STD 20.93, vs. 29.29, STD 16.60; mean increase 8.28), this increase is still not as significant as the difference between cell lines (mean drop 12.08) shown in Figure \ref{fig:cross}.}

\textcolor{black}{In 2024, Chandrasekaran et al. \cite{chandrasekaran2024three} established a correlation between chemical (compound) and genetic (gene-editing) perturbations based on morphological features extracted from CellProfiler. They identified that compound treatments can cause similar IF image appearances as gene editing, producing comparable morphological features. Subsequently, in this section, we investigated whether a model trained solely on compound-treated BF and IF images could accurately predict the IF images of gene-edited cells based on their BF images. Specifically, we wanted to see if the model could identify similar compound-gene pairs as observed in real images. Thus, we used CellProfiler to extract features from both real and synthetic images, aggregated the median values of each well, and used the t-SNE algorithm to compress the CellProfiler features into two dimensions. }

\begin{figure}[ht]
\centering
\includegraphics[width=\linewidth]{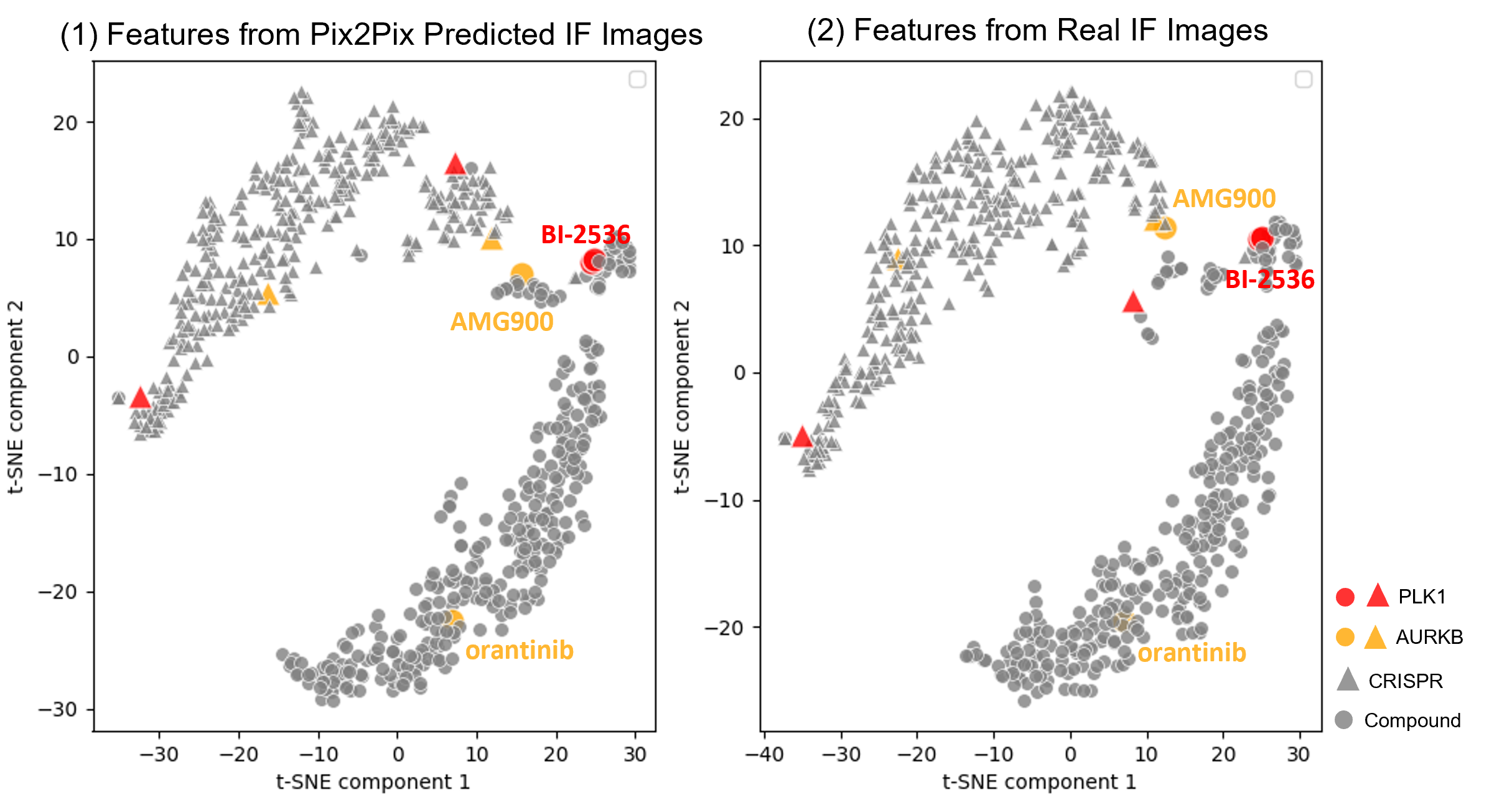}
\caption{T-SNE visualization of CellProfiler features extracted from synthetic A549 IF images generated by Pix2Pix (1) and from real A549 images (2). Triangles represent gene-edited images, and circles represent compound-treated images. Highlighted in both panels are the top positively correlated gene–compound matches identified by \cite{chandrasekaran2024three}. Both AMG900 (chemical) and orantinib (chemical) target AURKB. BI-2536 (chemical) is a PLK1 inhibitor compound, targeting PLK1 (gene). }
\label{fig:crispr}
\end{figure}

\textcolor{black}{In the study by Chandrasekaran et al., the top positively correlated gene–compound match in A549 cells is the PLK1 inhibitor compound BI-2536, which matched with CRISPR against PLK1, followed by the Aurora kinase inhibitor AMG900, which matched with CRISPR against AURKB. We highlighted these compound-CRISPR pairs in Figure \ref{fig:crispr}. We noticed that these highly similar pairs identified by real image features (red and orange dots in Figure \ref{fig:crispr} (2)) also exhibit high similarity in features extracted from synthetic images (red and orange dots in Figure \ref{fig:crispr} (1)). This underscores the generalizability of the Pix2Pix model to unseen perturbations and its ability to summarize biological phenomena as effectively as real images.}

\section{Discussion}
\subsection{Three Typical Artifacts in Synthetic IF Images}
In Figure \ref{fig:results}, we highlighted three types of artifacts that were identified when generative AI predicted IF images from BF images: 1) Blurring in UNet generated images (Figure \ref{fig:results}c): This artifact is prevalent in the UNet-predicted IF images through all channels. This is possibly attributed to the model's MSE-based optimisation loss function. The MSE loss encourages minimising pixel-wise differences, blurring fine details and leading to overly smoothed images. \textcolor{black}{Blurring can obscure critical cellular details, making it challenging to identify and analyse fine structures within the cells.} 2) Collapsed inner structure especially in the AGP channel: Some models fail to accurately replicate the structure, creating hallucinations or "collapsed biological structures." This hallucination is likely due to the model's inability to condition the synthesis procedure upon given BF conditions. \textcolor{black}{The presence of hallucinated or collapsed structures can mislead researchers about the actual actin, Golgi, plasma membrane (AGP) structures present in the sample, resulting in incorrect conclusions about cellular organization and function.} 3) Intensity shift and high background noises: In diffusion models, the predicted images often have significant background noise and intensity shifts. These artifacts can occur due to the nature of the diffusion process, where each step requires intensity clipping that may introduce noise and modify image brightness. \textcolor{black}{This intensity shift can complicate the quantification of fluorescence intensity, leading to potential errors in measuring protein expression levels, cell viability, or other quantitative metrics.}

We also observed that the Palette model inaccurately predicted cell locations and nuclei numbers, as illustrated in Figure \ref{fig:results}(3f). This loss of biological relevance may stem from neglecting certain conditions during the diffusion process. In many conditional diffusion models, the issue of ignoring conditions is commonly recognized \cite{tai2023revisiting,zhang2024enhancing}. Throughout the process, diffusion models may lose track of some elements of the prompts, resulting in the absence of objects in the final synthesised image. This issue arises as the training process of diffusion models is independent of the conditional factors \cite{ho2020denoising}. The conditioning strategy employed by the Palette model, which involves merely concatenating the BF images to guide the denoising process, appears to have more significant issues compared to the SPADE-diffusion approach. A possible solution to this issue is proposed by Cross-Zamirski et al. \cite{cross2023class}, wherein an additional condition of the compound target is added during the diffusion process. However, this requires prior knowledge of the type of perturbations which are not always readily available.

\subsection{Why Diffusion Models Underperform GANs?}
It had been suggested that Diffusion models should outperform GANs in \cite{dhariwal2021diffusion}. However, in our experiments, we discovered that diffusion models, including both Palette and Semantic Diffusion models, failed to predict IF images with high accuracy. Further, the performance drop was not the only problem identified in this task as, due to the denoising procedure, diffusion models took increased time to inference. In addition to decreased accuracy and increased inference time, another weakness of diffusion models is the potential skipping of input conditions, as shown in Figure \ref{fig:results} and discussed in section 5.1. 

The primary cause of these issues is the indirect optimisation target of diffusion models. The training objective focuses on reducing the MSE between the denoised and noised images using a UNet at each denoising step. However, this approach does not impose sufficient optimisation constraints on the final synthesised images after the entire denoising procedure. Additionally, even when the training loss has stabilised, the validation performance of the synthesised images can be inconsistent due to unstable factors during the denoising process.

\begin{figure}[ht]
\centering
\includegraphics[width=\linewidth]{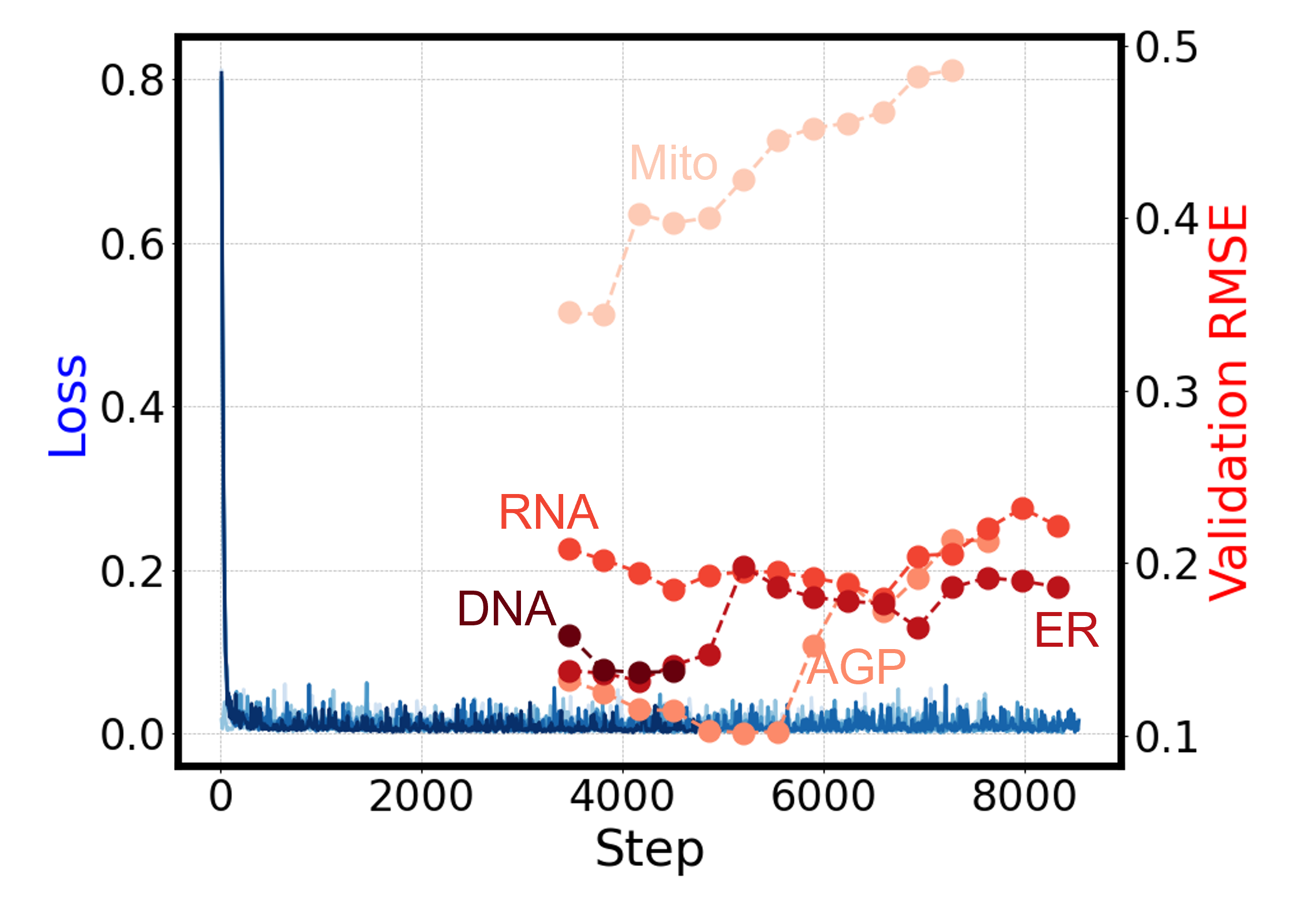}
\caption{The training loss curves of Palette for different channels and the validation RMSE. Even though the training loss converges, the RMSE loss exhibits significant perturbations that are caused by the indirect optimization target of the diffusion models.}
\label{fig:diffusion_loss}
\end{figure}

One way to mitigate these issues is to add additional guidance during the synthesis and denoising training procedure \cite{cross2023class}. However, this methodology is limited as it requires prior knowledge for the synthesis. Another method is to train each channel separately. Though this method can reduce the instability of diffusion models, it can not fully mitigate this issue, as is shown in Figure \ref{fig:diffusion_loss}. Therefore, we conclude that current diffusion models underperform compared to GAN models in the IF synthesis task. However, we did not evaluate latent diffusion models, which might help address these issues.

\subsection{Problematic MSE}

In this paper and most of the literature synthesising pseudo IF images,  Mean Squared Error (MSE) loss is used. MSE is defined mathematically as:

\[
\text{MSE} = \frac{1}{n} \sum_{i=1}^{n} (y_i - \hat{y}_i)^2
\]

where \( y_i \) is the ground truth pixel value and \( \hat{y}_i \) is the predicted pixel value. However, MSE loss is limited as an evaluation metric as it only considers the pixel value differences and aggregates the whole image regardless of background and signals. In a biological context, this may pose serious potential issues in downstream applications.

\begin{figure}[ht]
\centering
\includegraphics[width=\linewidth]{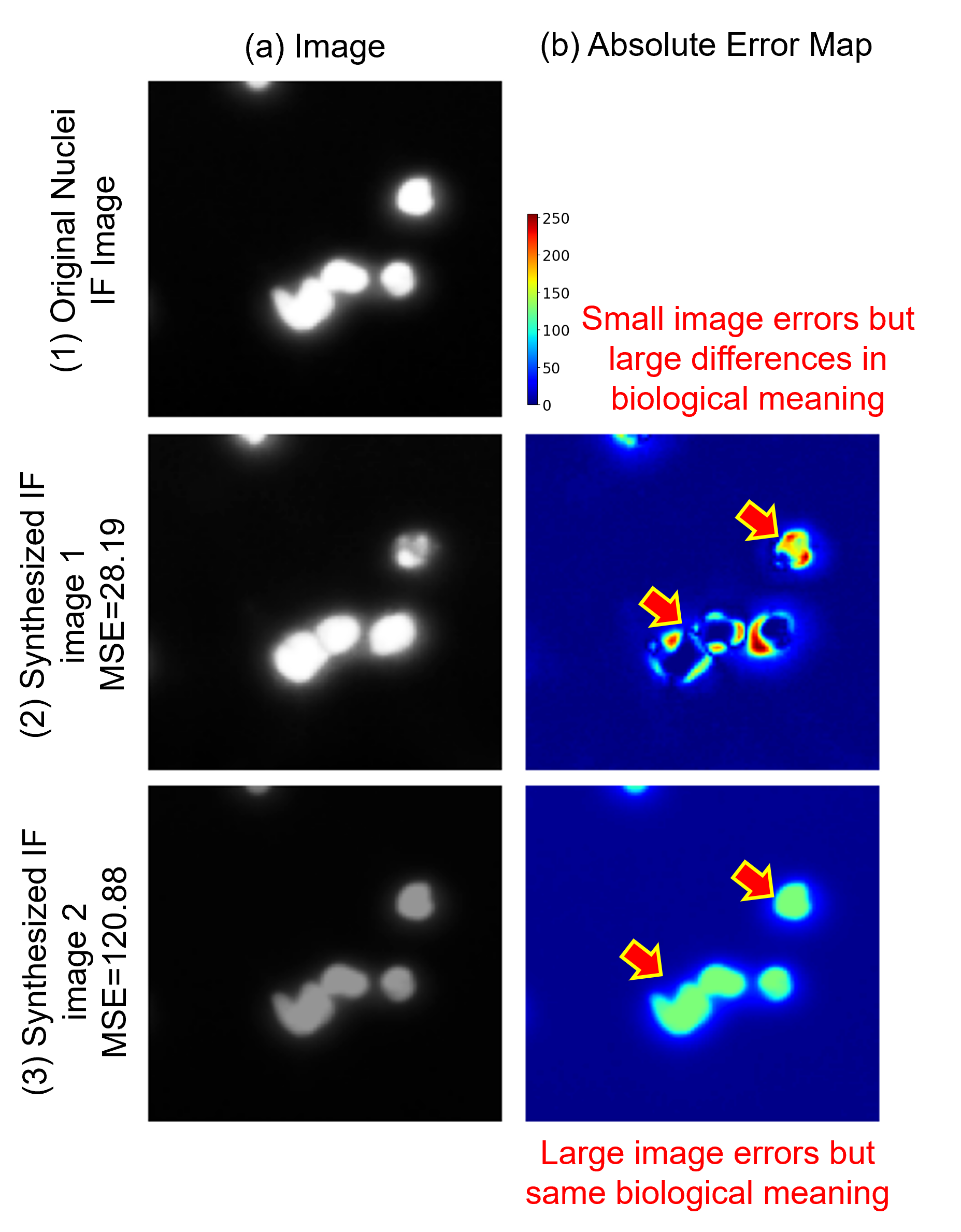}
\caption{MSE can be insufficient in IF evaluation scenarios. }
\label{fig:mse}
\end{figure}

 In Figure \ref{fig:mse}, we present two synthetic IF images (2,3) alongside their corresponding ground truth image (1). Figure \ref{fig:mse} (2a) was generated by the Pix2Pix network while Figure \ref{fig:mse} (3a) was produced by intensity reduction. As shown in Figure \ref{fig:mse} (2b), while Figure \ref{fig:mse} (2a) has a smaller image pixel intensity difference compare to the real image, and therefore a lower MSE loss, the cell morphology (in terms of nuclear size, shape and texture) is incorrectly predicted, leading to an inaccurate estimation of cell mortality, shown with two red arrows. On the other hand, Figure \ref{fig:mse} (3b) largely retains nuclear morphology despite large differences in stain intensity compared to the real image. Thus, while MSE, PSNR, and SSIM can roughly assess image intensity, they do not fully capture the biological relevance. As such, leveraging metrics that assess the biological significance of these synthetic images is crucial for robust model evaluation.

\subsection{Discrepancy in Feature Correlation and MoA Prediction Accuracy}
In Figure \ref{fig:cp_corr} and Table \ref{fig:results}, we observed that, although feature correlations of images generated by generative AI are low (mostly $<0.3$), the accuracy of single cell MoA predictions remains comparable to those made with real images. This is particularly evident in diffusion models that, despite having lower image quality evaluation scores, have greater accuracy for predicting MoA than real images.

\begin{figure}[h]
\centering
\includegraphics[width=\linewidth]{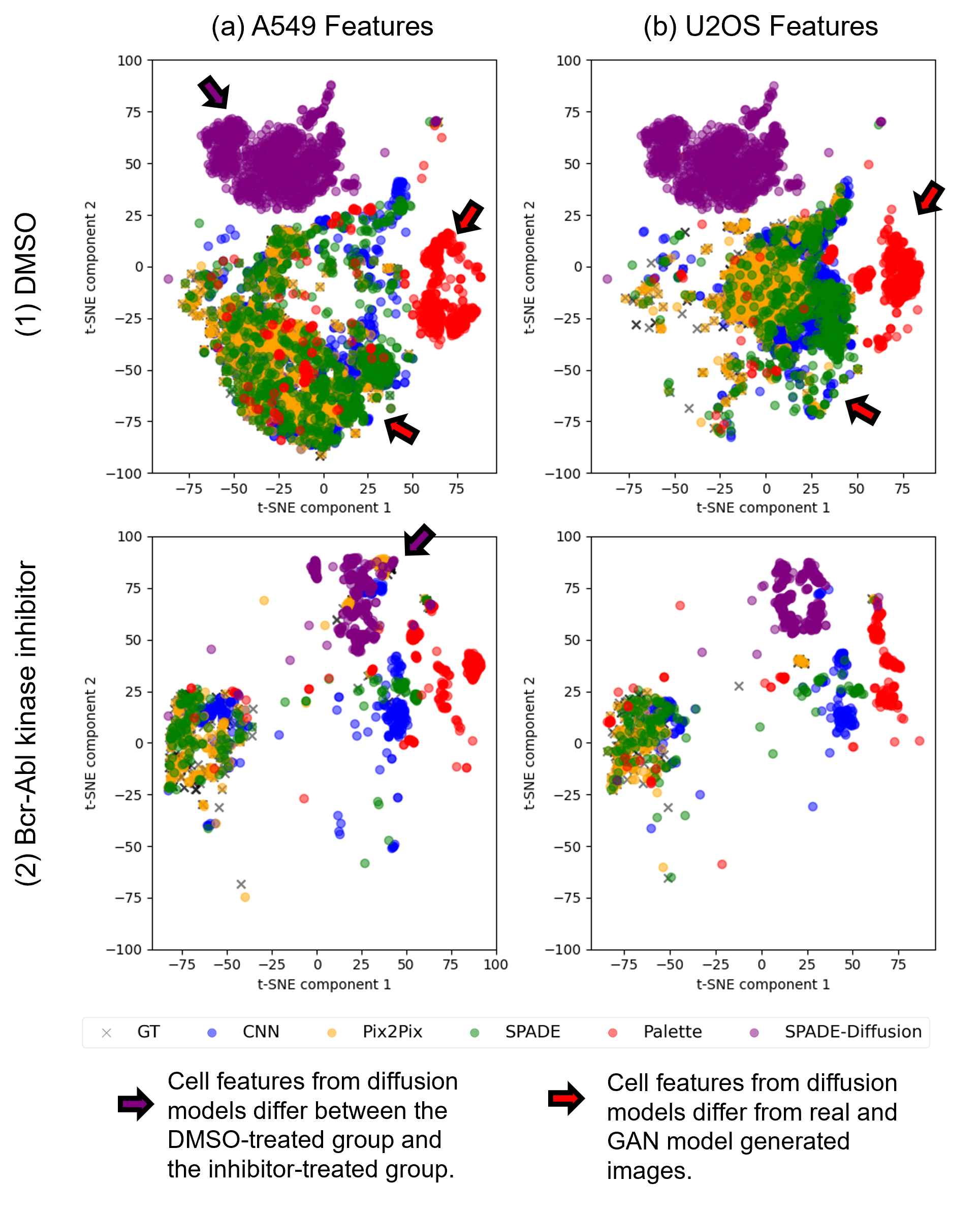}
\caption{The t-SNE maps showing the data distribution of cells. Each point represents a single cell identified from the IF images, coloured based on the methods used to generate the images.}
\label{fig:tsne}
\end{figure}

While diffusion models do not accurately capture the original, ground truth data distributions, they somehow capture the differences among cell phenotypes, allowing them to differentiate between the control group and the compound treatment group. This hypothesis is supported by plotting the feature distribution map using t-SNE, as shown in Figure \ref{fig:tsne}.

As illustrated, the single-cell features identified in UNet and GAN-based models closely follow the real feature manifold, which is consistent with the image quality results. In contrast, the features of cells generated by diffusion models occupy a different space and exhibit different characteristics. However, this does not imply that these out-of-real data points lack classification ability. Indeed, we observed a significant location difference between the DMSO group and the inhibitor-treated group in features generated by the diffusion models, as highlighted by the red and purple points in Figure \ref{fig:tsne}. \textcolor{black}{This indicates the potential of synthetic IF in predicting drug response and in high-content drug screening. The low feature correlation may harm the explainability of the classification model, but the synthetic images can still contribute valuable information for distinguishing between different treatment groups.}

\subsection{Can Generative AI Replace IF Staining Lab Work?}
At current performance, generative AI \textbf{cannot} directly replace IF staining wet lab work, but it does show significant potential, particularly if issues surrounding \textbf{generalisability} can be addressed. In this study, we produced synthetic images of high visual quality that achieved comparable performance in predicting MoA in single-cell states. Notably, a visual assessment from two experienced lab scientists (eyeballing) found that the IF images generated by the Pix2Pix model were indistinguishable from real ones.

Despite these promising results, we encountered a critical limitation regarding the generalisability of these data-intensive models. When applied to novel, unseen cell lines, there was a noticeable decline in performance (63.05\% in cell counting correlation and 9.72\% in SSIM). This issue underscores a practical challenge: if we have a comprehensive dataset sufficient to train a generative AI, the necessity of generative AI itself comes into question. The premise of generative models is to simulate scenarios where data is scarce or expensive to obtain, but if extensive data collection is already feasible, the benefits of using generative AI are diminished.

\textcolor{black}{The lack of generalisability also introduces another issue: the potential learning curve for biologists to obtain synthesis results. However, we argue that this learning curve correlates with the generalisability of the synthesis models. Running inferences is straightforward if a model is already trained. For example, image synthesis tools like Stable Diffusion \cite{stablediffusion} and Midjourney \cite{midjourney} are easily used by individuals without a computer science background because these models are highly generalisable and can directly generate synthetic images or responses.}

\textcolor{black}{Currently, we acknowledge that IF synthesis models lack this level of generalisability. For unseen cell lines, biologists would need to train their own models, which is more challenging. This process involves curating a dataset, acquiring training resources, and evaluating the models. Therefore, the generalisability of the models significantly impacts the practicality of virtual IF staining.}

\textcolor{black}{The limited generalisability of these models does not make them entirely obsolete. Firstly, synthetic IF images still hold significant potential in specific applications, such as longitudinal analysis and time-elapsed imaging, allowing for real-time, live imaging analysis. Secondly, the limitation can potentially be addressed by scaling up the model; Large generative AI models, such as StableDiffusion \cite{stablediffusion}, that have been trained on extensive, existing datasets have demonstrated generalisability to conditions the models had never seen before. It is likely this study does not approach the necessary size and scale required to produce such results.}



Furthermore, we observed that the features extracted from the synthetic images had a low correlation with those from real images. In addition, the MoA predictor, despite accuracy, relies on features that differ from those used with real data. These factors raise questions about the model's reliability and trustworthiness.

Another significant concern is the computational demand of these models, particularly the intensive GPU requirements for diffusion models. This computational burden presents a barrier to the widespread adoption and scalability of these generative AI approaches in practical laboratory settings.

\textcolor{black}{Last but not least, a significant concern regarding the use of generative AI in clinical settings is the ethical implications of using AI-generated images. The trustworthiness of these images can undermine the credibility of research papers that rely on synthesized data, since generated images are not "ground truths." The misuse of generative AI, including image manipulation and the creation of fake data, is another major concern.}

\subsection{General Generative AI for Biology: How Far Are We?}
In the previous discussion section, we examined how the lack of generalisability affects the utility of generative AI in IF image generation. One forward solution for increasing generalisability is scaling up the generative AI models with larger parameters and extensive datasets.

However, several challenges need to be overcome before scaling up. First is the lack of high-quality datasets. High-quality datasets covering a wide range of cell types and conditions are often limited. For example, StableDiffusion was initially trained on 2.3 billion images \cite{schuhmann2022laion}. In comparison, the entire cell painting repository \cite{cellpaintinggallery} contains fewer than 50 million images, where only 220k images were used in our paper. Acquiring a dataset comparable to those used in natural image synthesis models is impractical for a single lab due to the immense cost and effort required. However, this does not mean the task is impossible. Collaborative efforts among multiple institutions can pool resources and data, making it feasible to build comprehensive datasets.

Second is the issue of limited resources. Training generative AI models typically requires significant computational resources, including high-performance GPUs and large memory capacities, which can be a barrier for many labs. The financial and infrastructural demands for such resources are substantial, making it challenging for smaller institutions to participate in this advanced research.

\section{Conclusion}
This study is the first that has benchmarked the application of deep learning-based generators for synthesizing IF images from BF images to enhance the efficiency of high-content drug screening and cell mechanism of action analysis. We evaluated five different models based on three popular deep generator backbones, utilizing the open-sourced Cell Painting dataset.

Our key finding is that the results demonstrated that generative models can effectively synthesised IF images regarding their appearances. However, we conclude that generated IF images are not yet a replacement for IF staining due to limited generalisability, weak feature correlation, and the high GPU memory requirements for training these models.

\textcolor{black}{We suggest that future studies should adopt an evaluation framework that compares image quality, followed by an analysis of biological meanings using detailed feature correlations across multiple channels and subcellular parts, as well as MoA prediction accuracy using single cell features extracted from real and synthetic images.}

This paper not only serves as a comparative study to determine which model performs best but also proposes a comprehensive analysis pipeline to evaluate the efficacy of generators in IF image synthesis. While our findings highlight the potential of these generators, they also underscore the need for further research to address challenges such as model generalisability, feature relevance, and computational demands.

\bibliographystyle{cas-model2-names}

\bibliography{cas-refs}


\end{document}